\documentclass[twocolumn,showpacs,superscriptaddress,amssymb,10pt,pra,floatfix]{revtex4-1}

\usepackage{graphicx}
\usepackage{dcolumn}
\usepackage{bm}
\usepackage{epsfig}
\usepackage{xcolor}
\usepackage{longtable}
\usepackage{hyperref}
\hypersetup{hidelinks}
\usepackage{enumerate}
\usepackage{ulem}
\usepackage{amsmath}

\newcommand{\p}{\partial} 				
\newcommand{\e}{\mathrm{e}}				
\renewcommand{\i}{i}					
\renewcommand{\d}{\mathrm{d}}			
\renewcommand{\vec}{\bm}				

\newcommand{\Ai}{\mathrm{Ai}}	
\newcommand{\dK}{\d \kappa^4}		 		

\renewcommand{\emph}[1]{\textit{#1}}

\newcommand{\corr}[1]{\textcolor{blue}{#1}}
\renewcommand{\corr}[1]{{#1}}
\newcommand{\toproof}[1]{\textcolor{orange}{#1}}
\renewcommand{\toproof}[1]{{#1}}

\begin{document}

\preprint{}
%
%
%
%
\title{\corr{Impact of Earth's gravity on Gaussian beam propagation}\\ in hemispherical cavities}
%
%
%
%

\author{S.~Ulbricht}
\email{sebastian.ulbricht@ptb.de}
\affiliation{Physikalisch--Technische Bundesanstalt, D--38116 Braunschweig, Germany}
\affiliation{Technische Universit\"at Braunschweig, D--38106 Braunschweig, Germany}

\author{J.~Dickmann}
\email{johannes.dickmann@ptb.de}
\affiliation{Physikalisch--Technische Bundesanstalt, D--38116 Braunschweig, Germany}
\affiliation{Technische Universit\"at Braunschweig, D--38106 Braunschweig, Germany}
\affiliation{LENA Laboratory for Emerging Nanometrology, D--38106 Braunschweig, Germany}

\author{R.~A.~M\"uller}
\affiliation{Physikalisch--Technische Bundesanstalt, D--38116 Braunschweig, Germany}

\author{S.~Kroker}
\affiliation{Physikalisch--Technische Bundesanstalt, D--38116 Braunschweig, Germany}
\affiliation{Technische Universit\"at Braunschweig, D--38106 Braunschweig, Germany}
\affiliation{LENA Laboratory for Emerging Nanometrology, D--38106 Braunschweig, Germany}

\author{A.~Surzhykov}
\affiliation{Physikalisch--Technische Bundesanstalt, D--38116 Braunschweig, Germany}
\affiliation{Technische Universit\"at Braunschweig, D--38106 Braunschweig, Germany}
\affiliation{LENA Laboratory for Emerging Nanometrology, D--38106 Braunschweig, Germany}

\date{\today}

%
%
%
%

\begin{abstract} 
We theoretically investigate the influence of gravity on laser light in a hemispherical optical cavity, operating on Earth. The propagation of light in such a cavity is modeled by a Gaussian beam, affected by the Earth's gravitational field. On laboratory scale, this field is described by the spacetime of homogeneous gravity, known as Rindler spacetime. In that spacetime, the beam is bent downwards and acquires a height dependent phase shift. As a consequence the phase fronts of the laser light differ from those of a usual Gaussian beam.
Assuming that the initial beam enters the cavity along its symmetry axis,
these gravitational effects cause variations of the beam phase with every cavity round trip. 
Detailed calculations are performed to investigate how these phase variations depend on the beam parameters and the cavity setup. 
Moreover, we discuss the implications of our findings for cavity calibration techniques and cavity-based laser stabilization procedures. 

\end{abstract}
\maketitle
\section{Introduction} 

Laser interferometry and laser spectroscopy are today's most precise techniques to perform measurements at the frontiers of metrology \cite{Haensch2006}.
Due to that, in modern science they are indispensable to gain a better understanding of our world and to investigate nature.
The unprecendented precision of these techniques, would not have been possible without major advances in laser stabilization technology \cite{Ludlow2015,Kessler2012}. Therefore, the optical reference cavities used for this purpose are fundamental to the world's most precise measuring instruments like optical atomic clocks \cite{Ludlow2015, Ushijima2015, Huntemann2016}, atom interferometers \cite{Peters2001} and  next generation gravitational wave detectors \cite{Hogan2016, Kolkowitz2016}. Being essential to such a variety of devices, optical cavities help to answer open questions of physics, like the time-variation of fundamental constants \cite{Haensch2004}, the structure of the early universe \cite{openquest,lisa} and the nature of dark matter \corr{\cite{Stadnik2015,Stadnik2016,Derevianko2014,Geraci2019}}. 
%
%

The precision of cavity-based frequency measurements is limited by  a multitude of influences. 
The biggest of these influences are \corr{variations of the cavity's resonator length \cite{Kessler2012, Matei2017}, that} directly translate into an uncertainty of the cavity output frequency. 
Many sources of length variations, such as seismic vibrations and temperature fluctuations, can be suppressed in modern state-of-the-art cavities. In this case, the stability of the cavity is limited by the fundamental Brownian noise of the mirror coatings \cite{Kessler2012, Matei2017}. 
Using a cavity, which is fabricated from single-crystal silicon and cooled down to cryogenic temperatures, a  relative uncertainty of the resonator length and, thus, the output frequency, of $10^{-17}$ can be archived  \cite{Matei2017}.
Beyond that, future experiments at lower temperatures, utilizing crystalline coatings \cite{Cole2016} or meta mirrors \cite{Dickmann2018-2} are very promising to enhance the frequency stability of optical cavities by more than one order of magnitude. \looseness=-1
%

With further improvements of \corr{ laser frequency stability}, additional physical effects become relevant to a cavity setup.
Besides the well elaborated effects of quantum noise \cite{quantumnoise} and thermo-optic noise \cite{thermooptic}, one can expect that also the influence of gravity can not be neglected anymore. 
From Einstein's theory of general relativity it is known that the propagation of light is affected by gravity in the presence of heavy masses \cite{Einstein16,Edd19}.
Thus, also the light in a cavity is slightly deflected by the Earth's gravitational field  \corr{\cite{Raetzel2018,Rich19,Ulbricht2020}}.
In this work we, therefore, investigate theoretically, how \corr{the propagation of a laser beam is affected by gravitational light deflection with every cavity round trip.\linebreak} 
%
Our analysis starts in Sec.~\ref{sec:Rindler}, where we motivate Rindler spacetime as a model of the Earth's gravitational field on laboratory scale. 
The covariant Maxwell equations in this spacetime are used in Sec.~\ref{sec:maxwell} to obtain a wave equation, which accounts for the leading order gravitational effects on light propagation.
In Sec.~\ref{sec:IIb} and Sec.~\ref{sec:magic}, this wave equation is utilized to derive the vector potential for a gravitationally modified Gaussian beam. 
The obtained result is employed to study the propagation and reflection of light in a hemispherical cavity, \corr{consisting of a plane and a spherical mirror.}
In order to describe, how light evolves in such a cavity, in Sec.~\ref{sec:IIIa} we present a method to calculate the round trips of the gravitationally modified Gaussian beam iteratively.
The round trip calculation method is then used to study the phase of the beam at the \corr{plane cavity mirror} and to estimate the phase variations, caused by gravitational effects \corr{in Sec.~\ref{sec:IIIb}.}
\corr{In Sec.~\ref{sec:discussion} possible implications of our findings are discussed for a wide range of cavity setups, as it is used for laser frequency stabilization in Earth-based high precision experiments like atomic clocks and gravitational wave detectors.}
{The summary of the results and our conclusions are  given in  Sec.~\ref{sec:summary}.}
%

\section{Gaussian beams in a homogeneous gravitational field}\label{sec:propagation}

\subsection{The spacetime of homogeneous gravity} \label{sec:Rindler}
In the present work, we want to describe experiments with optical cavities in a laboratory on Earth.
In consequence, the equipment in such an experiment is affected by the Earth's gravitational field.
On usual laboratory scales, this gravitational field can be considered as homogeneous.   Additional effects, accounting for the Earth as a spherical body, can be neglected in a small region around the position of an observer. 
In the theory of general relativity, the co-moving frame of this observer is described by the spacetime of  homogeneous acceleration, i.e. Rindler spacetime \cite{Rind60, Rind66} with the line element
\begin{equation}
\d s^2 = g_{\mu\nu}\d x^\mu \d x^\nu =\left(1\corr{+}\frac{gz}{c^2}\right)^2\,\d(ct)^2-\d \vec{r}^2\,, \label{eqn:Metric}
\end{equation}
where $g$ is the module of $\vec{g}=- 9.81\, \mathrm{m}/\mathrm{s}^2\,\times\,\vec{e_z}$, which points into negative $z$-direction in the coordinate system $(ct,x,y,z)$.  Here,  $c$ is speed of light and $g_{\mu\nu}$ is the metric tensor with the sign-convention $(1,-1,-1,-1)$.  
Moreover, we use the Einstein notation, which means a sum is performed from 0 to 3 when paired Greek letters appear.
%
%

The line element (\ref{eqn:Metric}) describes merely flat Minkowski spacetime, but seen by an accelerated observer. Due to the non-geodesic motion of the observer it contains a $z$-dependent factor $(1\corr{+}gz/c^2)^2$, that accounts for the \emph{gravitational redshift} \cite{Rind60, Rind66}. 
Apparently, for vanishing acceleration the line element of Rindler spacetime reduces to the Minkowski line element. The same holds in the plane $z=0$, where flat Minkowski spacetime is reached asymptotically. 
%

\subsection{Light propagation in Rindler spacetime} \label{sec:maxwell}
As we know from general relativity, the properties of spacetime not only affect the motion of matter, but also the propagation of light.
This propagation mathematically is described by the wave equation.
In what follows, we will motivate the gravitational modifications of this wave equation in Rindler spacetime from first principles in a general relativistic framework \cite{Wald,Carrol}.
%
%

As usual in electrodynamics, the wave equation is obtained from the vacuum Maxwell equations, which in Rindler spacetime can be written in the covariant form
\begin{equation}
\nabla_\mu F^{\mu\nu}=0\,. \label{eqn:maxwell}
\end{equation} 
Those differential equations for the electromagnetic field strength tensor $F_{\mu\nu}=\p_\mu A_\nu-\p_\nu A_\mu$ are constructed from partial derivatives of the four-potential $A_{\mu}=(\,\Phi/c\,,\,\vec{A}\,)$, which contains the vector and scalar potentials $\vec{A}$ and $\Phi$, respectively. Moreover, in Eq.~(\ref{eqn:maxwell}) the covariant derivative $\nabla_\mu=\p_\mu+\Gamma^{\rho}_{\mu\rho}$  also carries information about Rindler spacetime, encoded in the Christoffel symbol\linebreak {$\Gamma^\rho_{\mu\rho}=g/c^2 \, \left(1\corr{+}gz/c^2\right)^{-1}\delta^{3}_\mu$}. By inserting the four-potential in Eq.~(\ref{eqn:maxwell}), the Maxwell equations in Rindler spacetime can be written as
\begin{eqnarray}
\nabla_\mu\nabla^\mu A^\nu=0\,, \label{eqn:covwave}
\end{eqnarray} 
where we assumed the Lorentz gauge condition  $\nabla_{\mu}A^{\mu}\!\!=\!0$.\linebreak
In Eq.~(\ref{eqn:covwave}) the equations of motion for the potentials $\Phi$ and $\vec{A}$ decouple. Since the scalar potential is fully determined by the Lorentz gauge condition \cite{Remark2}, we can restrict our discussion to the wave equation for the vector potential, which is given by
\begin{equation}
\frac{1}{c^2}\frac{\p^2}{\p t^2}\vec{A} -\vec{\mathcal{D}}^2\vec{A}=0 + \mathcal{O}(\epsilon^2)\,. \label{eqn:wave}
\end{equation}
Here $\vec{\mathcal{D}}=\left(1\corr{+}gz/c^2\right)\,\vec{\nabla}$ differs from the usual nabla operator $\vec{\nabla}=(\p_x,\p_y,\p_z)$ by the factor $\left(1\corr{+}gz/c^2\right)$, that now accounts for the gravitational redshift and \emph{light deflection}.
Moreover, we have considered effects of gravity only to linear order in the dimensionless parameter $\epsilon = gL/c^2$, which is in the range of $\epsilon \sim 10^{-18}\dots 10^{-13}$ for typical experiments. In these experiments the length scale $L$ ranges from $1\, \mathrm{cm}$ to $1\,\mathrm{km}$ and is small in comparison to the Earth radius, such that the gravitational field can be considered homogeneous.
%
%

In order to solve Eq.~(\ref{eqn:wave}), we assume paraxial light propagation. Restricting our discussion to the paraxial regime, the polarization vector $\vec{e}$ of the vector potential close to the axis of light propagation can be assumed to be coordinate independent \cite{Hecht2002}.
Under this assumption we make the ansatz
\begin{equation} 
\vec{A}(t,\vec{r})=T(t)\,X(x)\,Y(y)\,Z(z) \,\vec{e}\,, \label{eqn:ansatz}
\end{equation} 
where the functions $T$, $X$, $Y$ and $Z$ obey the linear ordinary differential equations
\begin{subequations}
\begin{eqnarray}
T''+\omega_0^2 T & = & 0\,, \label{eqn:Teq} \\
X''+k_x^2 X & = & 0 \,, \\
 Y''+k_y^2 Y & = & 0 \,, \label{eqn:Yeq} \\
 Z'' + 2 \gamma Z'+(k_z^2-\delta k_z^3 z) Z &  = & 0 \,.  \label{eqn:Airy}
\end{eqnarray}
\end{subequations}
In Eqs.~(\ref{eqn:Teq}) - (\ref{eqn:Yeq}), the constants $\omega_0$, $k_x$ and $k_y$ are arbitrary constants of separation, that determine $k_z^2$ in Eq.~(\ref{eqn:Airy}) by 
\begin{equation}
k_z^2  =  \omega_0^2/c^2 - k_x^2-k_y^2 \,, \label{eqn:Dispersion}
\end{equation}
The triplet $\vec{k}=(k_x,k_y,k_z)$ can be seen as a generalized wave vector and Eq.~(\ref{eqn:Dispersion}) as the corresponding dispersion relation. 
This picture is justified in the case of vanishing gravity, where $\vec{k}$ is a real valued vector. In contrast to that, in the presence of gravity, 
 $k_z^2$ from Eq.~(\ref{eqn:Dispersion}) can be also negative, to account for gravitational damping in $z$-direction. Moreover, an additional damping factor \linebreak $\gamma=g/2c^2$ and the height dependent deviation of the wave vector $\delta k_z  =  (2 g \omega_0^2/c^4)^{1/3}$ appears in  Eq.~(\ref{eqn:Airy}). As we will see in the next section, the latter gives rise to the Airy-kind nature of light propagation in $z$-direction.
%

\subsection{Solution to the wave equation up to $\mathcal{O}(\epsilon^2)$} \label{sec:IIb}
Above we derived the wave equation for the electromagnetic vector potential, that contains the influence of gravity on light propagation to leading order in $\epsilon=gL/c^2$. In what follows, we will solve this equation, using the ansatz (\ref{eqn:ansatz}). 

We find that Eqs.~(\ref{eqn:Teq})-(\ref{eqn:Yeq}) are solved by plane waves in the $(x,y)$-plane, while the solution to Eq.~(\ref{eqn:Airy}) is given by a damped Airy $\mathrm{Ai}$-function. Therefore, and due to the linearity of the wave equation (\ref{eqn:wave}), the vector potential $\vec{A}$ can be expressed as a linear combination of the basis functions
\begin{eqnarray}
\psi^{\vec{k}}_{\delta k_z}(t,\vec{r})= \frac{1}{\,2\pi \sqrt{\delta k_z}}\ \e^{\corr{-}\gamma z}\,\Ai\left[-\left(\frac{k_z}{\delta k_z}\right)^2\corr{+}\delta k_z z \right]\label{eqn:functions}\\
\times\, \e^{i k_x x} \e^{i k_y y} \e^{-i \omega_0 t} \,, \nonumber
\end{eqnarray}
that are defined for particular values of $k_x$, $k_y$ and $k_z^2$.
The normalization of these basis functions is chosen such that the integral over the full parameter space with the infinitesimal volume  $\dK=\d k_x \d k_y \d(k_z^2)$  gives
\begin{eqnarray}
\int\limits_{-\infty}^{\infty}\int\limits_{-\infty}^{\infty}\int\limits_{-\infty}^{\infty}& &\psi^{\vec{k}\ast}_{\delta k_z}(t,\vec{r}')\psi^{\vec{k}}_{\delta k_z}(t,\vec{r})\,\d k_x \d k_y \d(k_z^2) \hspace{7em} \nonumber \\
& &=[1\corr{-}\gamma (z+z')]\,\delta^{(3)}(\vec{r}-\vec{r}')+\mathcal{O}(\epsilon^2) \label{eqn:Delta}
 \\[0.5em]
& &=\hspace{2.1em}(1\corr{-}2\gamma z)\,\delta^{(3)}(\vec{r}-\vec{r}')+\mathcal{O}(\epsilon^2)\,\nonumber
\end{eqnarray}
and produces a spatial delta function up to a prefactor $(1+2\gamma z)$  (cf. Ref. \cite{Hunt81}). 
This prefactor, arising from the damping terms $\e^{\corr{-}\gamma z}=1\corr{-}\gamma z +\mathcal{O}(\epsilon^2)$ cancels the redshift factor in the coordinate invariant infinitesimal space volume $(1\corr{+}2\gamma z)\d x\d y \d z$, such that the spatial integral over (\ref{eqn:Delta}) is one. 
The relation (\ref{eqn:Delta}) implies that the set of basis functions (\ref{eqn:functions}) is complete and can be used to express the vector potential $\vec{A}$ as superposition of the $\psi^{\vec{k}}_{\delta k_z}(t,\vec{r})$. In the limit of vanishing gravitation $g=0$, the set of basis functions $\{\psi^{\vec{k}}_{\delta k_z\to 0}(t,\vec{r})\}$ becomes a plane wave basis $\{\e^{\i( \vec{k}\cdot\vec{r}-\omega_0 t)}\}$, which describes light propagation in the spacetime of an inertial observer.
This can be retraced to the asymptotic behavior of the Airy $\mathrm{Ai}$-function \cite{Vallee}.
In this limit, moreover, the normalization integral (\ref{eqn:Delta}) becomes an integration over the standard Fourier space $\d k_x \d k_y \d k_z$, which produces a spatial delta function.
%
%

\vspace{-1em}
\subsection{Propagation of a Gaussian beam in Rindler spacetime}\label{sec:magic}
In what follows, we want to discuss the findings of the previous section for the particular case of a horizontally propagating Gaussian beam. This beam is a suitable model for laser light, as it is used in many experimental devices, such as laser stabilization cavities \cite{horizontalcavity} and gravitational wave detectors \cite{Saulson}. Choosing the $x$-axis as the propagation axis of the beam, we define the boundary condition of the Gaussian beam at $x=0$ by
\begin{equation}
\vec{A}(t,0,y,z)= \left(1\corr{-}\frac{gz}{2c^2}\right)\frac{\vec{A}_0}{2\pi b_0^2}\,\e^{-\frac{z^2+y^2}{2b_0^2}}\e^{-\i \omega_0 t}\,, \label{eqn:boundary}
\end{equation}
where $b_0$ is the beam waist. Moreover, we require the boundary condition to share the  redshift behavior $\e^{\corr{-}\gamma z}=(1\corr{-}gz/2c^2)+\mathcal{O}(\epsilon^2)$ of the basis functions (\ref{eqn:functions}) as the fundamental solution to the wave equation (\ref{eqn:wave}). In order to express the boundary condition in terms of the basis functions (\ref{eqn:functions}), we make use of the delta function and Eq.~(\ref{eqn:Delta}) to obtain
\begin{eqnarray}
& &\hspace{-1em}\vec{A}(t,0,y,z)\nonumber\\
&=&\int\vec{A}(t,0,y',z')\delta^{(3)}(\vec{r}-\vec{r}')\d{\vec{r}'}^3\label{eqn:trafo}\\
&=&\int\!\!\!\!\!\int\!\!\!\vec{A}(t,0,y',z')\psi^{\vec{k}\ast}_{\delta k_z}(t,\vec{r}')\psi^{\vec{k}}_{\delta k_z}(t,\vec{r})\,(1\corr{+}2\gamma z')\,\dK\d {\vec{r}'}^3\nonumber\\
&=&\int\!\!\tilde{\vec{A}}^b_{\vec{k}}(t)\psi^{\vec{k}}_{\delta k_z}(t,\vec{r})\,\dK \,.\nonumber
\end{eqnarray}
Here the transformation of the vector potential at the boundary 
\begin{eqnarray}
\tilde{\vec{A}}^b_{\vec{k}}(t)
&=&\vec{A}_0\exp\left[{-\frac{1}{2}(k_z^2+k_y^2)b_0^2}\right]\delta(k_x)\hspace{4em}\label{eqn:Ak}\\
& &\times\frac{1}{\sqrt{\delta k_z}}\Ai\left[-\left(\frac{k_z}{\delta k_z}\right)^2+\frac{1}{4}(\delta k_z b_0)^4\right]\,.\nonumber
\end{eqnarray}
is obtained by evaluating the spatial integral in Eq.~(\ref{eqn:trafo}), \cite{Remark1}. 
The first line of expression (\ref{eqn:Ak}) is the Fourier transformation of a spatial Gaussian profile, as it is expected in the case of vanishing gravity $g=0$. 
The gravitational corrections in Eq.~(\ref{eqn:Ak}) are introduced by the Airy $\mathrm{Ai}$-function in the second line.
%
%

\begin{figure}[t]
	\centering
	\includegraphics[scale=0.21]{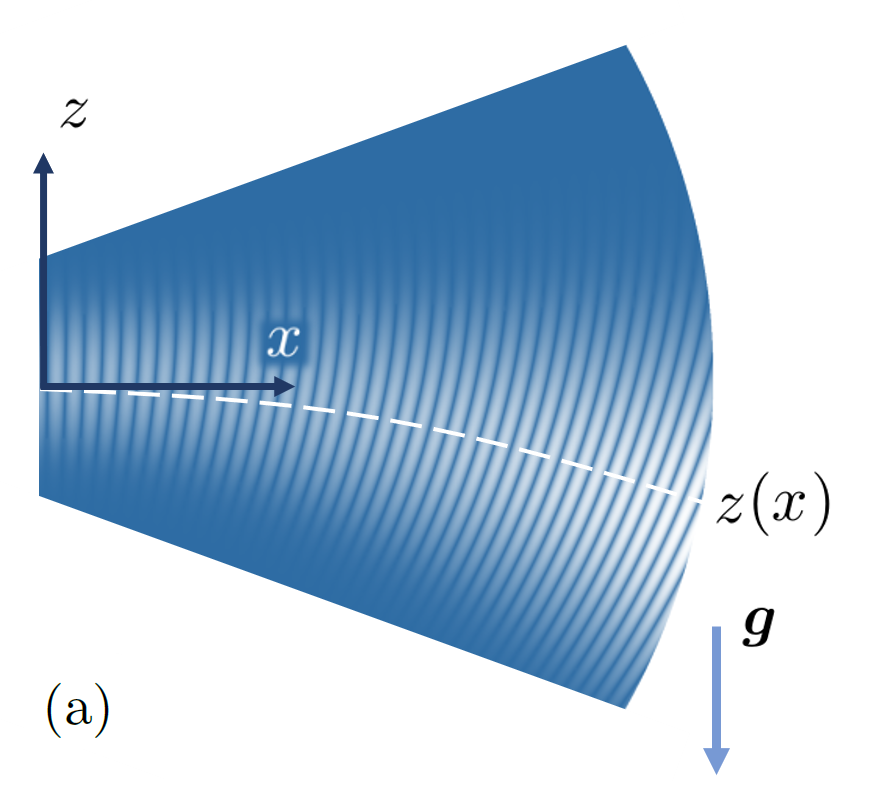}$\!\!\!$\includegraphics[scale=0.21]{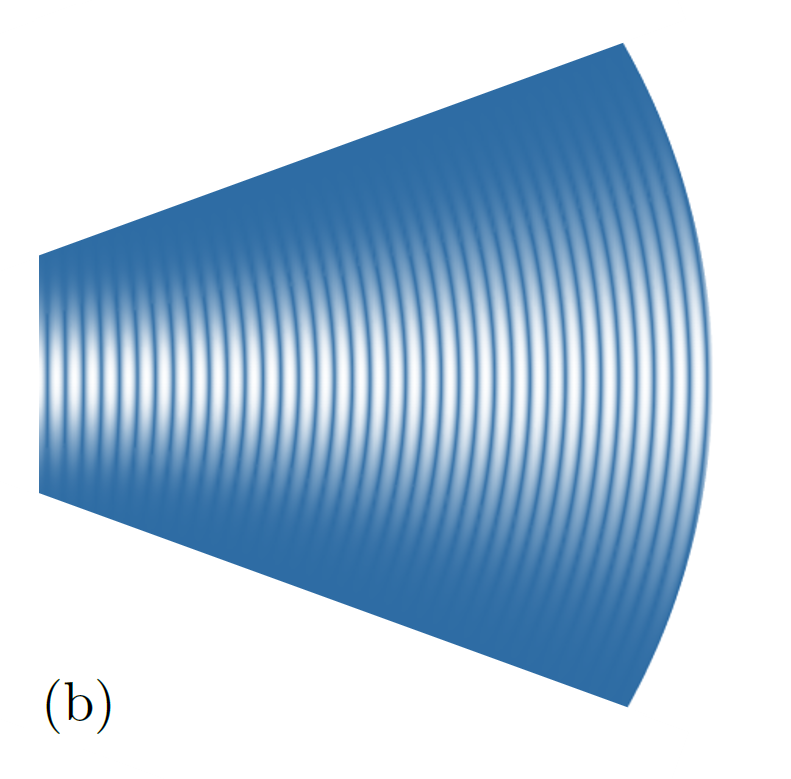}
	\caption{Schematic picture of the Gaussian beam propagation in a hemispherical cavity in the presence of a homogeneous gravitational field (a) and without gravity (b). \corr{In an scenario, where the symmetry axis of the initial beam coincides with the cavity symmetry axis,} in the presence of gravity, the wave fronts \corr{deviate from} the surface of the spherical mirror. Moreover, the intensity maximum follows the line $z(x)=-gx^2/2c^2$ in the gravitational affected case.}
	\label{fig:phase propagation}
\end{figure}

Eqs.~(\ref{eqn:trafo}) and (\ref{eqn:Ak}) only describe
the vector potential $\vec{A}(t,0,y,z)$ at the boundary $x=0$. 
However, we aim to find the vector potential of the Gaussian beam $\vec{A}(\vec{r},t)$ in the entire coordinate space and for all times. Therefore, we recall that the vector potential has to obey the wave equation (\ref{eqn:wave}) and the generalized dispersion relation (\ref{eqn:Dispersion}), everywhere in space.
This requirement can be fully fulfilled by the replacement
\begin{eqnarray}
\delta(k_x) \quad&\to&\quad \delta\left(k_x - \sqrt{\omega_0^2/c^2 -k_y^2-k_z^2}\right)\label{eqn:delta}\,
\end{eqnarray}
in the Eqs.~(\ref{eqn:trafo}) and (\ref{eqn:Ak}).
Moreover, the resulting replacement in the transformation of the vector potential $\tilde{\vec{A}}^b_{\vec{k}}(t)\to \tilde{\vec{A}}_{\vec{k}}(t)$ still has the proper boundary condition (\ref{eqn:boundary}). Since we know that the solution to the wave equation is uniquely determined by the boundary condition (e.g.~\cite{Dirichlet}), we obtain the vector potential in the full coordinate space by 
\begin{eqnarray}
\vec{A}(t,\vec{r}) =\int\!\!\tilde{\vec{A}}_{\vec{k}}(t)\psi^{\vec{k}}_{\delta k_z}(t,\vec{r})\,\dK\,. \label{eqn:newA}
\end{eqnarray}
Having made the substitution (\ref{eqn:delta}) in Eqs.~(\ref{eqn:trafo}) and (\ref{eqn:Ak}) and making use of the paraxial approximation $\sqrt{\omega_0^2/c^2 -k_y^2-k_z^2} \approx \omega_0/c-c(k_z^2+k_y^2)/2\omega_0$ for\linebreak $k_x\gg k_y,k_z$, we can perform the $k_x$-integral over the delta function in Eq.~(\ref{eqn:newA}) to obtain
\begin{eqnarray}
& &\vec{A}(t,\vec{r})= \vec{A}_0 \e^{\corr{-}\gamma z}\, \e^{\i \frac{\omega_0}{c}(x-ct)}\nonumber\\ 
& & \times\,\, \frac{1}{\sqrt{2\pi}}\int\limits_{-\infty}^{\infty} \e^{-\frac{1}{2}k_y^2 \mathcal{B}(\mu)} \e^{\i k_y y'}\d k_y \label{eqn:anoterA}\\
& &\times\,\, \frac{1}{\sqrt{2\pi }\,\delta k_z}\int\limits_{-\infty}^{\infty} \,\e^{-\frac{1}{2}k_z^2 \mathcal{B}(\mu)}
\,\Ai\left[ -\left(\frac{k_z}{\delta k_z}\right)^2+ \frac{1}{4}(\delta k_z b_0)^4 \right]\nonumber\\
& &\hspace{9.6em}\times \,\Ai\left[ -\left(\frac{k_z}{\delta k_z}\right)^2\!\corr{+}\, \delta k_z z' \right]\,\d (k_z^2)\,. \nonumber
\end{eqnarray}
Here we introduced the complex quantity \linebreak $\mathcal{B}(\mu)=b_0^2(1+\i \mu)$,with $\mu=x/x_R$ being the propagation distance in units of the Rayleigh length $x_R=\omega_0 \,b_0^2/c$ \cite{Meschede}.
In order to further investigate the vector potential $\vec{A}(t,\vec{r})$, we perform the Fourier transformation in the second line of Eq.~(\ref{eqn:anoterA}) and employ Eq.~(\ref{eqn:Delta}) to calculate the $k_z^2$-integral (cf. Ref.~\cite{Vallee}).
We obtain
\begin{eqnarray}
\vec{A}(t,\vec{r})=\left(1\corr{-}\frac{gz}{2c^2}\right)\vec{A}^{\mathrm{G}}(t,\vec{r})\e^{S_g(\vec{r})} + \mathcal{O}(\epsilon^2)\,\,, \label{eqn:fullbeam}
\end{eqnarray}
as a product of an unperturbed Gaussian beam
\begin{eqnarray}
\vec{A}^{\mathrm{G}}(t,\vec{r})=\frac{\vec{A}_0 }{2\pi b_0^2(1+\mu^2)}\exp\left[-\frac{y^2+z^2}{2b_0^2(1+\mu^2)}\right]\hspace{5em}\label{eqn:GaussianBeam}\\
\times\,\exp\left[{\i \omega_0 x/c - \i \omega_0 t+\frac{\i (y^2+z^2)\mu}{2b_0^2(1+\mu^2)}-\i\arctan(\mu)}\right]\nonumber
\end{eqnarray}
and an exponential function with the complex argument
\begin{eqnarray}
S_g(\vec{r})=\corr{-}\frac{ g\omega_0^2 b_0^2 z}{2c^4(1+\mu^2)}\left[\mu^2+\i (2 \mu+ \mu^3)\right]\,.  \label{eqn:Correction}
\end{eqnarray}
This complex exponent accounts for the leading order effects of a homogeneous gravitational field and results in two major modifications of the Gaussian beam:
On the one hand, the real part of $S_g(\vec{r})$ leads to a bending downwards of the beam's intensity profile, so that the intensity maximum of the beam follows the line $z(x)=-gx^2/2c^2$, see Fig.\ref{fig:phase propagation}. On the other hand, the imaginary part of $S_g(\vec{r})$ gives rise to a $z$-dependent gravitational phase shift $\phi_g(x\gg x_R)\corr{=-} g\omega_0/c^3\,\, z \,x $, that grows while the beam propagates along the $x$-axis. 
%
%


While in \cite{Ulbricht2020} we concentrated on the implications of the gravitational modifications (\ref{eqn:Correction}) for the intensity profile of the Gaussian beam, in the present work we will focus on their consequences for its phase.
In order to do this, we formally rewrite Eq.~(\ref{eqn:fullbeam}) in the form $\vec{A}(\vec{r},t)=|\vec{A}(\vec{r},t)| e^{\i\phi(\vec{r},t)}\,\,\vec{e}$, where the phase of the gravitationally modified Gaussian beam, \corr{up to the Gouy phase $\arctan(\mu)$}, is given by
\corr{
\begin{subequations}
\begin{eqnarray}
\phi(\vec{r},t)&=&\frac{\omega_0}{c}\!\left(\! x+\frac{y^2+z^2-gz(2x^2_R+x^2)/c^2}{2b_0^2(x+x_R^2/x)}-ct\!\right)\label{eqn:phase1}\qquad\\
&\approx &\frac{\omega_0}{c}\!\left(R(\vec{r})-\frac{x_R^2}{x}-ct\right) + \mathcal{O}(\epsilon^2)\label{eqn:phase2}
\end{eqnarray}
\end{subequations}}
\corr{with $R(\vec{r})$ defined as}
\begin{equation}
\corr{R(\vec{r})^2=\left(\!x+\frac{x_R^2}{x}\right)^2+y^2+\left(\!z-\frac{g}{c^2}\left(x_R^2+\frac{x^2}{2}\right)\right)^2\!. \label{eqn:phase3}}
\end{equation}
$\!\!$Here we recognized that Eq.~(\ref{eqn:phase1}) is the paraxial approximation of Eq.~(\ref{eqn:phase2}). 
The \corr{phase properties of the gravitationally affected Gaussian beam} can be illustrated in terms of \corr{equal-time} phase fronts $\phi(\vec{r},t_0)\corr{=\omega_0(L/c- t_0)}=\mathrm{const}$. \corr{
	In order to obey this requirement, $R(\vec{r})$ from (\ref{eqn:phase2}) has to be substituted by
	\begin{eqnarray}
	R(\vec{r})=L+\frac{x_R^2}{x}\,.
	\end{eqnarray}
	By this replacement, Eq. (\ref{eqn:phase3}) parameterizes spheres of radius  $R(x=L)=L+x_R^2/L$, with their focus at a horizontal distance $x_0(L)=-x_R^2/L$. 
	This resembles the well known phase properties of a Gaussian beam, that starts as a plane wave at $L=0$ and becomes a spherical wave for $L\gg x_R$ \cite{Meschede}.
	In the presence of gravity, however, the beam additionally is affected by gravitational light deflection, such that the spherical phase fronts in Eq. (\ref{eqn:phase3}) are shifted by a vertical distance 
	\begin{eqnarray}
	z_0(L)=\frac{g}{c^2}\left(x_R^2+\frac{L^2}{2}\right) \label{eqn:shift}
	\end{eqnarray}
	with respect to the height of the initial Gaussian beam at $x=0$.
This shift into the positive $z$-direction becomes more pronounced with increasing propagation distance and opposes the bending downwards of the intensity profile as shown in Fig. 1. With this, we have identified the leading gravitational effects on Gaussian beam propagation. In what follows,  we will analyze their consequences for optical setups in a laboratory and for Earth-based cavity experiments in particular.}
%

\clearpage
\section{Hemispherical Cavities}\label{sec:cavities}
In the previous section we developed a formalism to describe a Gaussian beam, that is affected by a homogeneous gravitational field. The results of this analysis can be used to discuss a wide range of applications in Earth-based optical experiments.
In what follows, we consider the particular example of beam propagation in a \corr{cavity resonator, as it is utilized in various laser stabilization procedures}
%
%

\corr{As we know, in an optical setup, the phase fronts of a beam have to coincide with the mirror surfaces, in order to transpose the beam into it self by reflection. In the special case of a Gaussian beam, therefore, a mirror of radius $R(L)$ has to be placed at a distance $x=L$ from a plane mirror at $x=0$, in order to construct a phase matched resonator. Eq. (\ref{eqn:shift}) implies that, in the presence of gravity the beam would have to enter this \emph{hemispherical cavity} at a vertical position $-z_0(L)$ below the cavity symmetry axis.
Since $z_0(L)\sim L^2$, the relevance of this effect increases for longer cavity devices. Moreover, the needed vertical shift grows with $x_R^2 \sim b_0^4$, such that it becomes more pronounced for large beam waists.
Especially close to the ultimate case $b_0\to\infty$, where the beam approaches the plane wave limit and the cavity becomes co-planar, the effect of light deflection can not be compensated by varying the vertical beam position.
These findings indicate that, depending on the properties of the experimental setup, the influence of the Earth's gravitational field has to be taken into account in the cavity calibration procedure.}
%

\corr{In the following sections we discuss a scenario, where the symmetry axis of the initial beam coincides with the cavity symmetry axis. 
In this scenario, the properties of the beam slightly deviate between two round trips in the resonator. The resulting phase deviations at the plane mirror will be analyzed to quantify the impact of gravity on the beam propagation a hemispherical cavity.}
%

\subsection{Round trip calculation method}  \label{sec:IIIa}
In order to consider gravitational effects \corr{in a hemispherical cavity}, we first have to describe, how the beam properties evolve with every reflection \corr{at the cavity mirrors}.
For that purpose, let us follow the beam on its first round trip, as displayed in Fig.~\ref{FIG03}:
\begin{figure}[t]
	\includegraphics[scale=0.27]{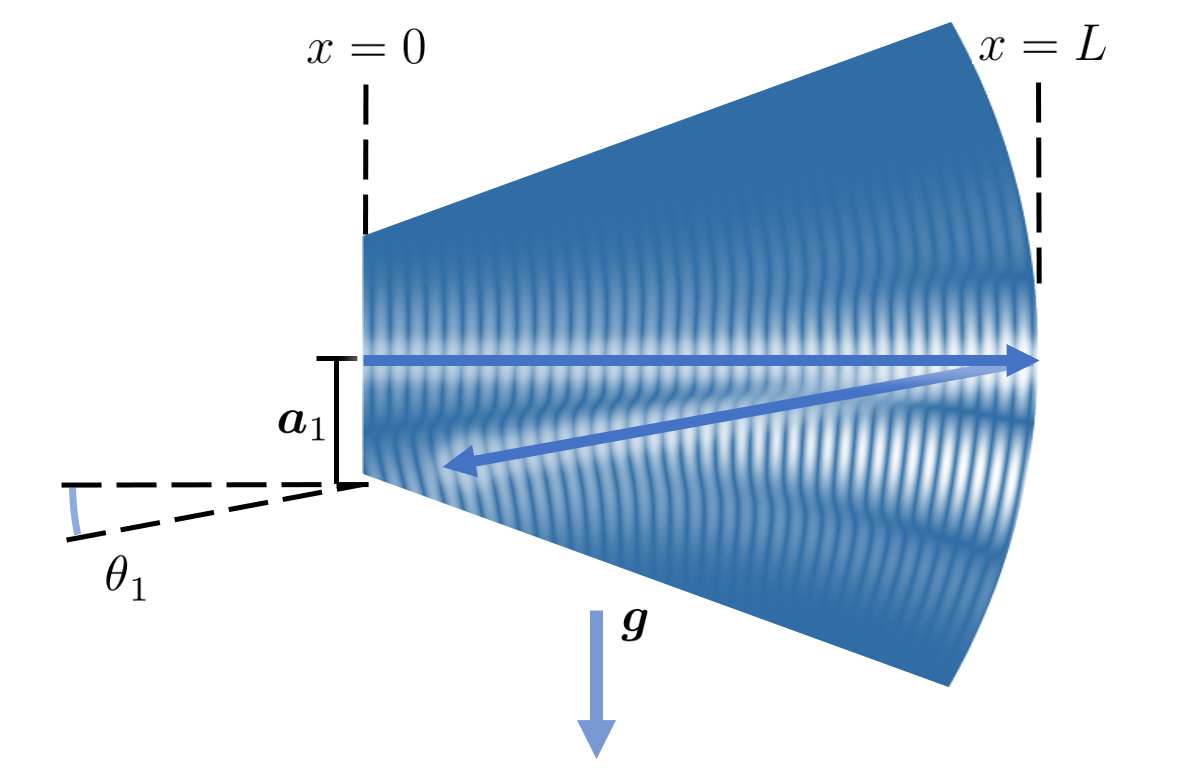}
	\caption{First reflection of the gravitational modified Gaussian beam in a hemispherical cavity. Due to the homogeneous gravitational field, the modified Gaussian beam is no eigen-mode of the cavity. In result the beam axis of the initial beam are not reflected into itself by the spherical mirror.  This deviation of the reflected beam axis from the initial one is modeled by the angle $\theta_1$ and the shifting parameter $a_1$. } \label{FIG03}
\end{figure}
%

After entering the cavity at $x=0$, the gravitational modified Gaussian beam propagates towards the spherical mirror. 
As discussed in Sec.~\ref{sec:magic}, on this way, its behavior deviates from common Gaussian beam propagation.
That discrepancy becomes crucial, when the light reaches the spherical mirror, which is designed to reflect the unperturbed beam (\ref{eqn:GaussianBeam}).
In result, it is not transposed into itself, but reflected under a non-zero angle. After reflection, the beam again is affected by gravity and, therefore, has to obey the wave equation (\ref{eqn:wave}). 
Following these considerations, we can express the reflected light, utilizing the solution (\ref{eqn:fullbeam}), but with a slightly changed direction of propagation. 
Thus, we assume that the vector potential after reflection can be written as 
\begin{equation}
\vec{A}^{(1)}(\vec{r},t)=\vec{A}^{(0)}(\vec{r}_1,t)\,, \label{eqn:rtm}
\end{equation}
where $\vec{A}^{(0)}(\vec{r},t)=\vec{A}(\vec{r},t)$ denotes the initial beam, which coincides with the Gaussian beam (\ref{eqn:fullbeam}).
The change of the propagation direction
\begin{equation}
 \vec{r}_1={\hat R}(\theta_1) \cdot \vec{r} + a_1\vec{e}_z\,  \label{eqn:series}
\end{equation}
is determined by the angle $\theta_1$ and position $a_1$ under which the reflected beam hits the plane mirror at $x=0$ in the end of the first round trip. Since these orientation coefficients are of linear order in $\epsilon= g L/c^2$, the propagation axis of the beam after reflection keeps horizontal and, hence, remains perpendicular to $\vec{g}$ to the desired order $\mathcal{O}(\epsilon^2)$.
%

The parameters $\theta_1$ and $a_1$ can be found in an elegant way by comparing the phase fronts of the initial beam $\vec{A}^{(0)}(\vec{r},t)$ and the reflected beam $\vec{A}^{(1)}(\vec{r},t)$ at the position of the spherical mirror.
To do that, we calculate the wave vectors $\vec{k}^{(0)}=\vec{\nabla}\phi(\vec{r},t)$ and $\vec{k}^{(1)}=-\vec{\nabla}\phi(\vec{r}_1,t)$, which at the spherical mirror have to obey the relation
\begin{equation}
\vec{k}^{(1)} = \vec{k}^{(0)} - 2 (\vec{k}^{(0)}\cdot\vec{n})\,\vec{n}\,.
\end{equation}
Here $\vec{n}\sim \left.\vec{\nabla}\phi(\vec{r},t)\right|_{g=0}$ is the normal vector of the spherical mirror, which is normalized to one (cf. Ref. \cite{Hecht2002}).
%

In the example above, we have considered a single round trip of light in the cavity. In order to start the next round trip we only have to change $x\to -x$ in $\vec{A}^{(1)}(\vec{r},t)$ due to the reflection at the plane mirror and repeat the procedure explained above. That way, we obtain a series 
$\vec{r}_n={\hat R}(\theta_n) \cdot \vec{r} + a_n\vec{e}_z$, which characterizes the orientation of the beam $\vec{A}^{(n)}(\vec{r},t)=\vec{A}(\vec{r}_n,t)$ in every round trip.
%

\subsection{Properties of the beam orientation coefficients} \label{sec:III2}
In the previous section, we developed a method to obtain the orientation of the gravitational modified Gaussian beam for every round trip in the hemispherical cavity.
While the presented method can be used to analyze a wide range of specific cavity setups, characterized by the beam waist $b_0$, cavity lengths $L$ and wavelengths $\lambda$ of the used laser light, in this section we restrict our discussion to two limiting cases.
\begin{figure}[b]
	\includegraphics[scale=0.66]{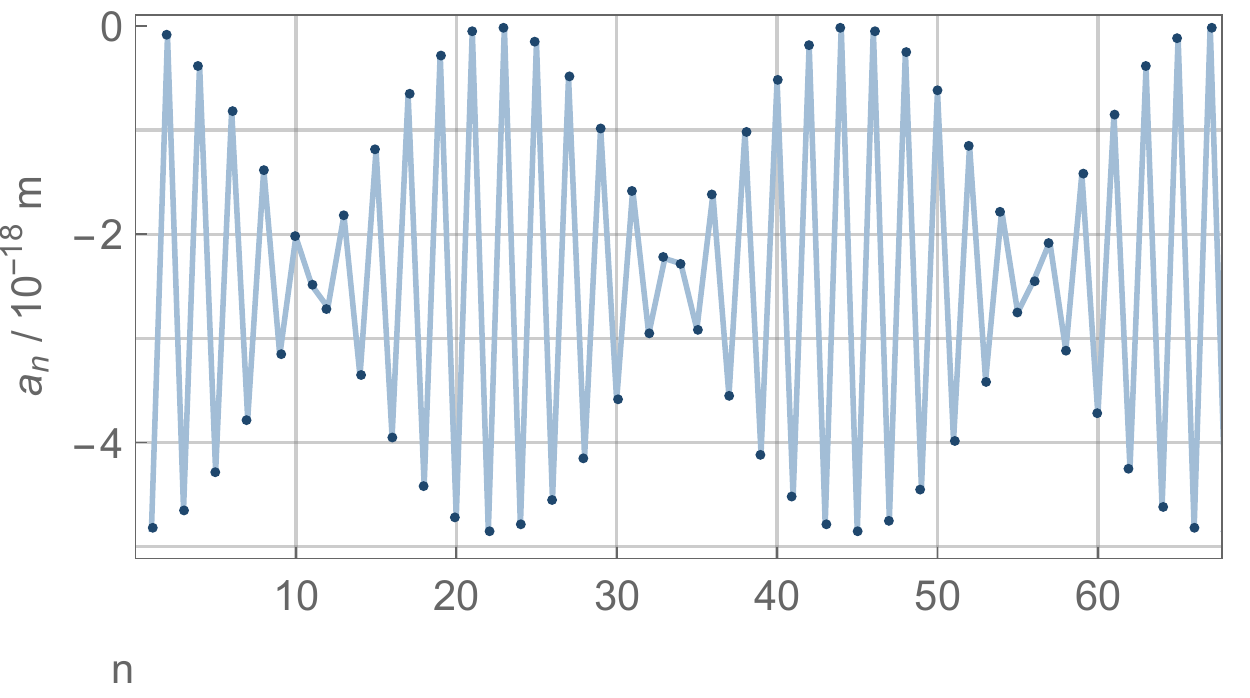}\\
	\includegraphics[scale=0.648]{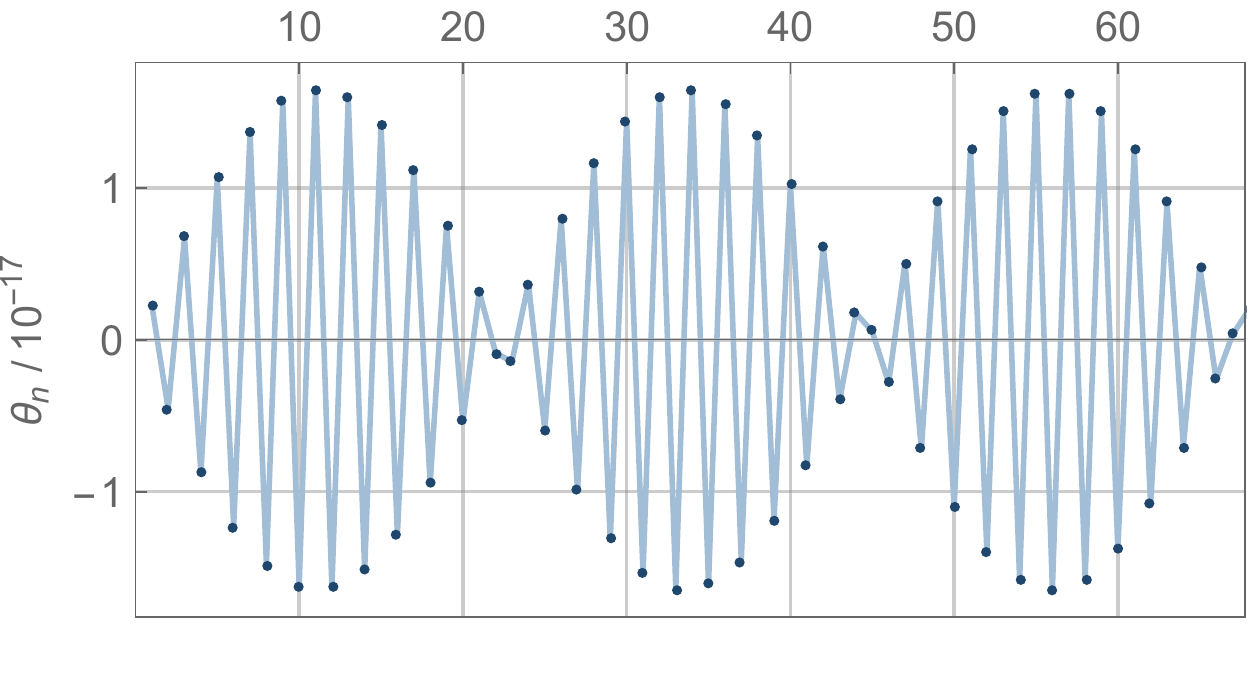}\,\,\\
	\caption{Oscillations of the beam orientation coefficients $a_n$ and $\theta_n$ in the parameter regime $b_0\ll\sqrt{L\lambda}$ for a 21 cm cavity with a beam waist of $b_0=50\, \mu$m for light with $\lambda=1064$ nm. Due to the strong focusing effect of the spherical mirror, the series alternates with $(-1)^{n}$.\\ \\[0.1em]} \label{FIG08}
\end{figure}
First, we consider the case of small beam waists $b_0\ll\sqrt{L\lambda}$, in which the phase properties of the Gaussian beam resemble the properties of a spherical wave.
In this regime, the orientation coefficients are given by
\begin{subequations}
\begin{eqnarray}
a_n
&=& -\frac{gL^2}{2c^2}\left[1+(-1)^{n+1}\cos\!\left(\frac{4\pi b_0^2}{L\lambda}\,n \right)\right]\,\,\,\,\,\, \label{eqn:ansmall}\\[0.5em]
\theta_n
&=&\frac{gL}{c^2}(-1)^{n+1}\!\frac{L\lambda}{4\pi b_0^2}\sin\!\left(\frac{4\pi b_0^2}{L\lambda}\,n \right) \,, \label{eqn:tnsmall}
\end{eqnarray}
\end{subequations}
as displayed in Fig.~\ref{FIG08} for a beam with $b_0=50\,\mu \mathrm{m}$ and $\lambda=1064\,\mathrm{nm}$ in a $L=21\,\mathrm{cm}$ cavity. As seen from the figure and Eqs.~(\ref{eqn:ansmall}) and  (\ref{eqn:tnsmall}), the evolution of both parameters is described by an alternating series, which is modulated by a periodic envelope with the frequency $\pi b_0/L\lambda$. The parameters $a_n$ and $\theta_n$ oscillate around the mean values $\langle\theta\rangle=0$ and $\langle a\rangle=-gL^2/2c^2$, respectively.
The periodicity of the parameters can be explained due to the interplay of two effects: Since the beam is bent downwards by gravity, it tends to leave its initial axis of propagation. 
However, the further the beam removes from its initial propagation axis, the more it is reflected back into the direction of this axis by the focusing effect of the spherical mirror. In result, for small, but finite beam waists $b_0$, the hemispherical cavity acts as a stable resonator.  However, that is not the case for a vanishing beam waist $b_0\to 0$, where the orientation coefficients become $a_n(b_0\to 0)=\frac{gL^2}{2c^2}\{(-1)^{n}-1\} $ and\linebreak $\theta_n(b_0\to 0)= \frac{gL}{c^2}\,(-1)^{n+1}\,n$. 
\begin{figure}[b]
	\includegraphics[scale=0.66]{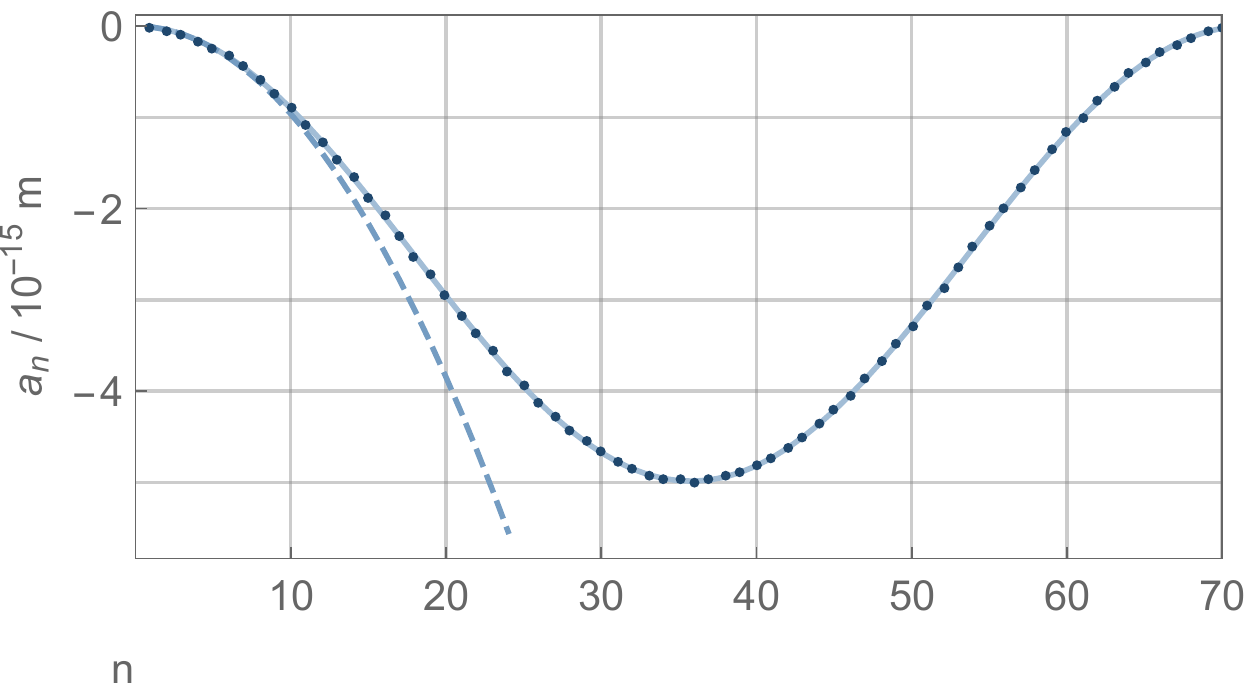}
	\includegraphics[scale=0.66]{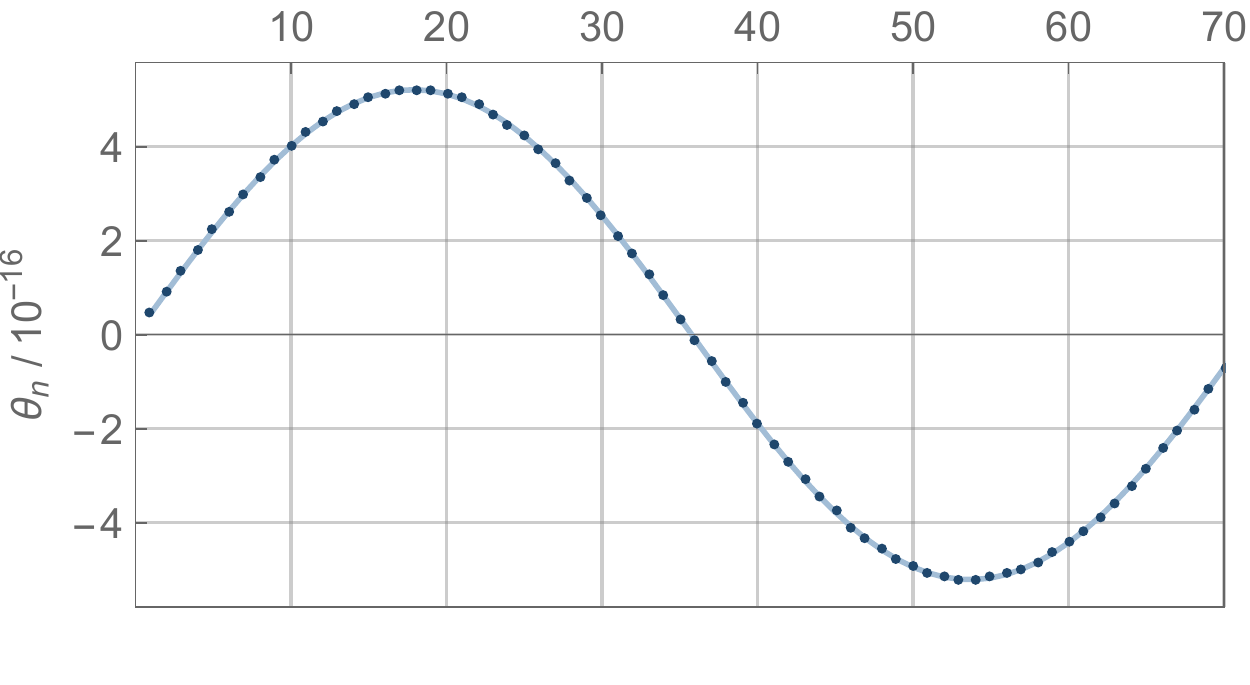}\\
	\caption{ Oscillations of the beam orientation coefficients $a_n$ and $\theta_n$ in the parameter regime $b_0\gg\sqrt{L\lambda}$ for a 21 cm cavity with a beam waist of $b_0=900\, \mu$m for light with $\lambda=1064$ nm. The shift $a_n$ initially follows the trajectory of free fall $-\frac{g}{2}(2Ln)^2$ (dashed), before the weak focusing effect of the spherical mirror leads the beam back to the initial axis of propagation.} \label{FIG08b}
\end{figure}
As the second limiting case, we consider a gravitational modified Gaussian beam with a large beam waist $b_0\gg\sqrt{L\lambda}$, where the orientation coefficients with every round trip are given by 
\begin{subequations}
\begin{eqnarray}
a_n
&=& \frac{gL^2}{c^2}\left(\frac{2\pi b_0^2}{L\lambda}\right)^2\left[\cos\!\left(\frac{L\lambda}{\pi b_0^2} \,n\!\right)-1\right]\label{eqn:anbig}\,\,\,\,\,\\[0.5em]
\theta_n
&=&\frac{gL}{c^2}\!\frac{2\pi b_0^2}{L\lambda}\sin\!\left(\frac{L\lambda}{\pi b_0^2} \,n\!\right) \,. \label{eqn:tnbig}
\end{eqnarray}
\end{subequations}
In Fig.~\ref{FIG08b} this is shown for a beam with $b_0=900\,\mu \mathrm{m}$\linebreak and $\lambda=1064\,\,\mathrm{nm}$ in a $L=21\,\mathrm{cm}$ cavity.
From the figure and the Eqs.~(\ref{eqn:anbig}) and (\ref{eqn:tnbig}) it can be seen that both parameters again oscillate around their mean values, which now read $\langle\theta\rangle=0$ and\linebreak $\langle a\rangle=-gL^2/c^2\times \left(2\pi b_0^2/L\lambda\right)^2\corr{=-gx_R^2/c^2}$, respectively. The frequency of these oscillations is given by $L\lambda/\pi b_0^2$, what, in contrast to the case of small $b_0$, becomes smaller with increasing beam waists. Therefore, also in the regime of large, but finite $b_0$ the cavity acts as a stable resonator. However, the cavity becomes unstable in the ultimate case $b_0\to \infty$, when the orientation coefficients reduce to $a_n(b_0\to \infty)=-\frac{g}{2}\frac{(2Ln)^2}{c^2}$ and $\theta_n(b_0\to \infty)=2 \frac{gL}{c^2}\,n$. In that limit, the phase properties of the Gaussian beam approach those of a plane wave. Hence, the resonator becomes a Fabry-Pérot cavity with a plane mirror at $x=L$. This mirror has no focusing effect and does not prevent the beam from falling down. Therefore, after $n$ round trips and a propagation distance of $2Ln$ in the cavity, the beam finds itself at a height $a_n=-\frac{g}{2c^2} (2nL)^2$, as expected for a light ray, bent by Earth's gravity in the limit of geometrical optics \cite{Rich19,Wald}. 
%

\subsection{Phase \corr{variations at the plane cavity mirror}} \label{sec:IIIb}
Previously, we laid down a theory to describe the propagation of light in a hemispherical cavity, that is affected by the Earth's gravitational field. Below, we apply this theory to  cavities with the particular resonator lengths of $L=$  21 cm, 30 cm and 50 cm, which currently represent the world's most stable optical frequency references \cite{Kessler2012,Didier2019,Keller2014,Nakagawa1994} and, therefore, are indispensable for the improvement of high precision 
instruments, such as optical atomic clocks \cite{Ludlow2015, Ushijima2015, Huntemann2016} and gravitational wave detectors \cite{Hogan2016, Kolkowitz2016}.
In what follows, we will focus on the gravitational effects on the output frequency of these devices. 
One can expect that the stability of the output frequency is affected by the gravitational distortion of the light propagation, discussed in Sec.~\ref{sec:propagation}. Due to this distortion, the beam hits the plane mirror, \corr{which commonly is used as the cavity output}, at slightly different positions in each round trip. Hence, also the phase of the vector potential $\vec{A}^{(n)}(\vec{r},t)$ slightly varies with every reflection in the cavity. We can calculate that effect by using the theory, developed in Sec.~\ref{sec:IIIa} and the summation of the contributions from all round trips \cite{Ismail2016}, which are characterized by the orientation coefficients $a_n$ and $\theta_n$. 
Since in real experiments the mirrors have a finite \corr{power} transmittance $T$, a fraction of light leaves the cavity with every reflection and only an effective number of round trips $N$ has to be considered. This number is in the range of the cavity finesse \corr{$\mathcal{F}\approx \pi/T$} \cite{Dawkins2007}, which is above $10^{5}$ for recent experiments \cite{Kessler2012,Matei2017,Ludlow2007}.
%

Having discussed qualitatively, how gravitational \corr{light deflection} affects the phase \corr{of the Gaussian beam} at the \corr{at the plane cavity mirror}, we are ready to perform this analysis quantitatively.
We start with Eq.~(\ref{eqn:phase1}), which directly relates the trajectory of the beam and its phase.
From that expression we can define the \corr{local phase deviations} at the \corr{plane cavity mirror}
\begin{figure}[t]
	\includegraphics[scale=0.65]{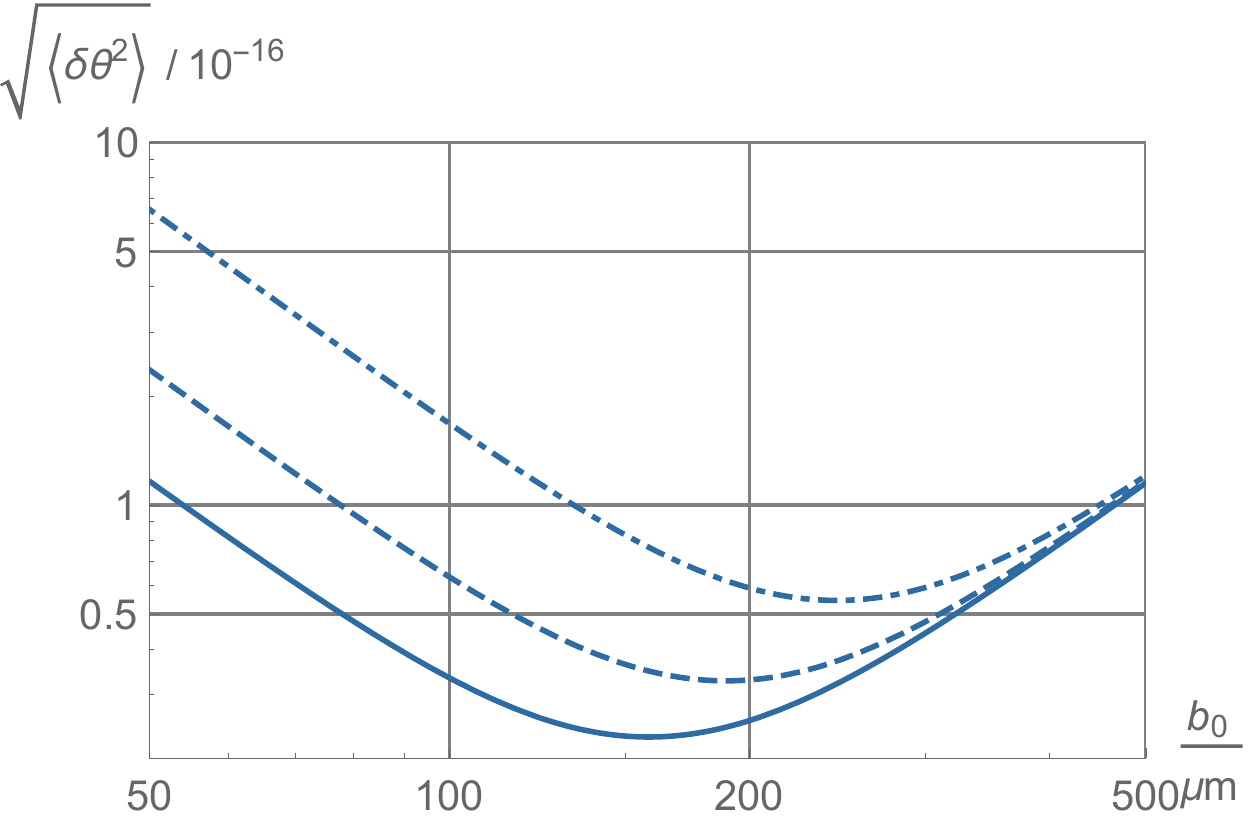}\\[1em]
	\caption{Averaged deviation of the angle $\sqrt{\langle \delta\theta^2 \rangle}$  for light with a wave length of $\lambda=1064$ nm as a function of the beam waist $b_0$ for 21 cm cavities (solid), 30 cm cavities (dashed) and 50 cm cavities (dot-dashed).} \label{FIG05}
\end{figure}
\begin{eqnarray}
\delta \phi(\vec{r}) &=& \left.\sqrt{\left\langle\left(\phi( \vec{r},t)-\phi( \vec{r}_n,t)\right)^2\right\rangle}\right|_{x=0}  \label{eqn:phasediff1}
\end{eqnarray}
as the averaged difference between the phase of the initial Gaussian beam and the phase of the beam on every round trip.
Since the beams orientation $\vec{r}_n$ only slightly differs from $\vec{r}$, we can formally rewrite $\vec{r}_n=\vec{r}+\delta \vec{r}_n+\mathcal{O}(\epsilon^2)$, where the structure of $\delta \vec{r}_n=(\,\theta_n z\,,\,0\,,\,a_n-\theta_n x \,)$ follows from Eq.~(\ref{eqn:series}). That way, we can expand Eq.~(\ref{eqn:phasediff1}) around $\vec{r}$, to obtain 
\begin{eqnarray}
\delta \phi(\vec{r}) &=& \left.\sqrt{\left\langle\left(\vec{\nabla}\phi(\vec{r},t)\cdot\delta\vec{r}_n\right)^2\right\rangle}\right|_{x=0} \label{eqn:phasediff2}\,.
\end{eqnarray}
Due to the fact that $\vec{r}_n$  already is of linear order in $\epsilon=gL/c^2$, we can evaluate the phase gradient of the unperturbed Gaussian beam (\ref{eqn:GaussianBeam}) at $x=0$. At this position, the phase fronts of the beam are those of plane waves, such that $\left.\vec{\nabla}\phi(\vec{r},t)\right|_{x=0}=(\omega_0/ c\,,\,0\,,\,0\,)+\mathcal{O}(\epsilon)$. This simplifies Eq.~(\ref{eqn:phasediff2}), which now reads 
\begin{eqnarray}
\delta \phi(\vec{r}) 
&=& \frac{z\omega_0}{c} \sqrt{\langle \delta\theta^2 \rangle}+ \mathcal{O}(\epsilon^2) \label{eqn:phasediff}\,.
\end{eqnarray}
As seen from the above expression, $\delta \phi(\vec{r})$ becomes zero in the case of vanishing gravity, since the phase of the beam remains unchanged with every round trip. In contrast, the presence of the gravitational field leads to non-zero phase \corr{deviations, which are} proportional to the light frequency $\omega_0$ and the \corr{averaged deviation} of the tilting angle $\sqrt{\langle \delta\theta^2 \rangle}$, which can be calculated as the 
root mean square value of the series $\theta_n$:
\begin{eqnarray}
 \sqrt{\langle \delta\theta^2 \rangle} \,\,&=&\sqrt{\frac{1}{N}\sum\limits_{n=1}^N (\theta_n)^2}\,\quad. \label{eqn:tas}
\end{eqnarray}
In this formula, we took into account that the periods of oscillations in $\theta_n$ are much smaller, than the effective number of round trips $N\sim\mathcal F$, such that no relative weighting of the contributions from every round trip has to be considered. Eq.~(\ref{eqn:tas}) was used to calculate $\sqrt{\langle \delta\theta^2 \rangle}$ for light with a wave length of $\lambda=1064$ nm as a function of $b_0$ for cavities with a resonator length of $L=$ 21 cm, 30 cm and 50 cm, see Fig.~\ref{FIG05}. As can be seen there, the averaged deviation of the angle increases with $L$ and is rather sensitive to the beam waist $b_0$.
%

Eq.~(\ref{eqn:phasediff}) describes the \corr{local phase deviations} at some particular point \corr{of the plane cavity mirror, located at $x=0$}. In order to calculate the variations of the phase \corr{over the entire mirror}, we have to weight it with the intensity distribution of the Gaussian beam is obtained:
\begin{eqnarray}
\sqrt{\langle \delta\phi^2 \rangle} &=&\sqrt{\frac{1}{\pi b_0^2} \int\,\e^{-\frac{z^2+y^2}{b_0^2}}\delta \phi(\vec{r})^2\,\d y \, d z}\nonumber\\
&=& \frac{b_0 \omega_0}{2c}  \sqrt{\langle \delta\theta^2 \rangle}\,.
\label{eq:DeltaPhi}
\end{eqnarray}
\corr{In what follows, we will analyze, how $\sqrt{\langle \delta\phi^2 \rangle}$ as measure for the gravitationally induced phase variations at the plane mirror, commonly used as the cavity output, depend on the cavity parameters and discuss its possible implications for Earth-based cavity experiments.}
%

\section{Discussion} \label{sec:discussion}
\begin{figure}[t]
	\includegraphics[scale=0.54]{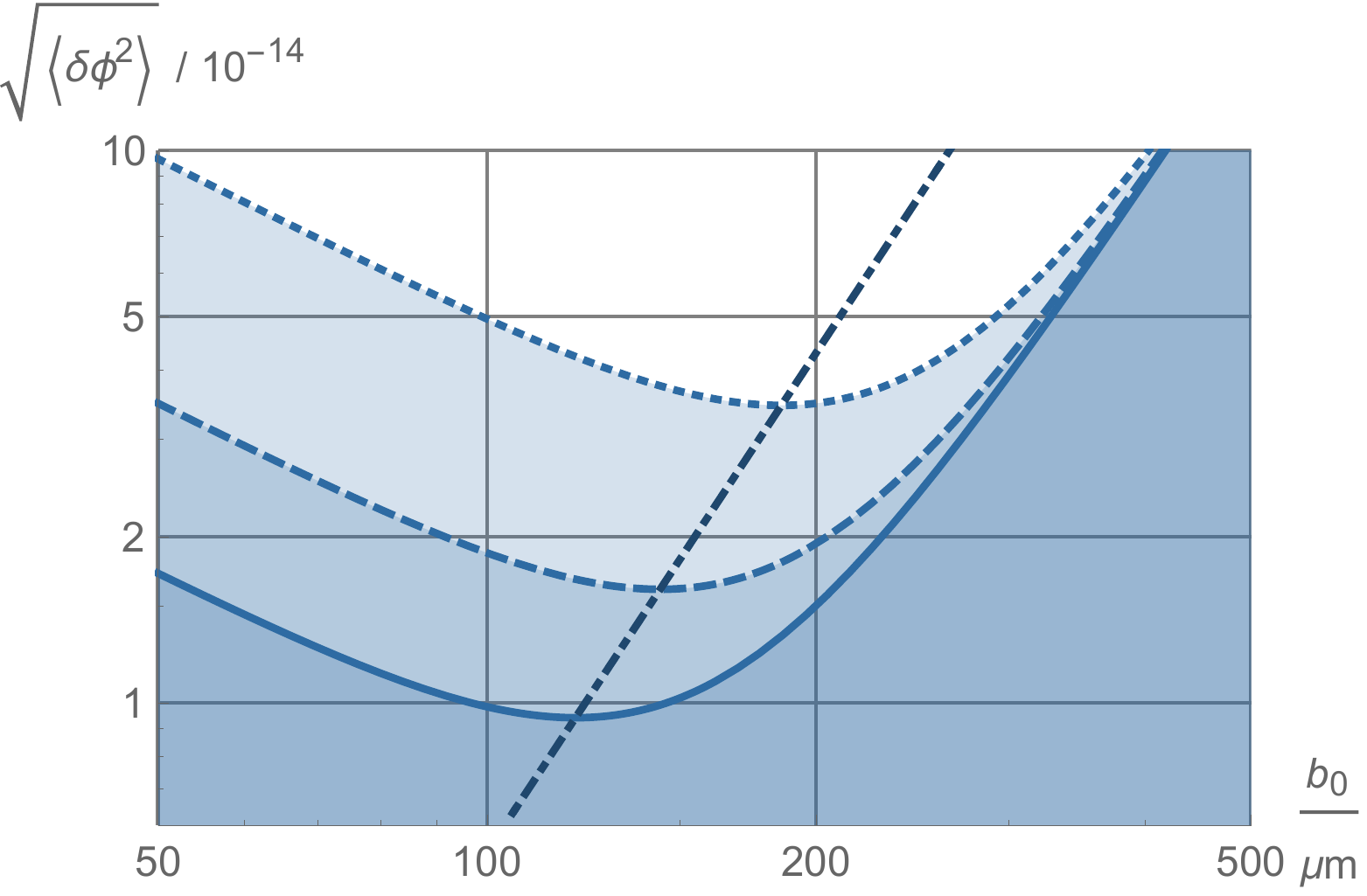}
	\hspace{3em}
	\caption{ \corr{Phase variations $\sqrt{\langle \delta\phi^2 \rangle}$ at the plane mirror of} a hemispherical cavity for light with $\lambda=1064$ nm, depending on the beam waist $b_0$, for 21 cm cavities (solid), 30 cm cavities (dashed) and 50 cm cavities (dotted). 
		In the logarithmic axis scales of this plot all three curves have the same shape. They are only shifted along the dashed straight line of optimum beam waist (dot-dashed). 
	} \label{FIG07}
\end{figure}

Above, we developed a theory to analyze the propagation of a Gaussian beam in a hemispherical laser cavity, affected by the Earth's gravitational field.
This theory was used to investigate, how gravity influences the
propagation direction of a beam, \corr{which initially enters the cavity along its symmetry axis.} 
Applying the methods, presented in Sec.~\ref{sec:cavities} for a wide range of cavity settings, we find that \corr{in this scenario the phase of the Gaussian beam at the plane cavity mirror differs with every round trip. The root mean square of these deviations, as a quantitative measure of the gravitational phase variations,  is given by}
\begin{equation}
\corr{\sqrt{\langle \delta\phi^2 \rangle}} = \frac{g}{\sqrt{2} c^2}\corr{\frac{2\pi L}{\lambda}}\left(\frac{L\lambda}{8\pi b_0}+\frac{\pi b_0^3}{L\lambda}\right) \label{eqn:freqstab}\,,
\end{equation}
depending on the characteristic lengths scales $L$, $\lambda$ and $b_0$ of the cavity system. In order to illustrate the dependence of $\corr{\sqrt{\langle \delta\phi^2 \rangle}}$   on theses parameters, in Fig.~\ref{FIG07} \corr{the phase variations are} displayed for light with $\lambda=1064\,\mathrm{nm}$ and cavity lengths of 21 cm, 30 cm and 50 cm.
As seen from that figure, \corr{the gravitational phase variations} strongly depend on the beams waist.
For small $b_0<100\,\mu\mathrm{m}$, for example, $\corr{\sqrt{\langle \delta\phi^2 \rangle}}$ is proportional to $1/b_0$. In contrast, for large values of $b_0>300\,\mu\mathrm{m}$, it grows with $b_0^3$. This behavior can be traced back to the $b_0$-dependence of the orientation coefficients 
(\ref{eqn:tnsmall}) and (\ref{eqn:tnbig}) and the definition (\ref{eqn:tas}).
%

The increase of \corr{the phase variations} for small and large beam waists, implies the existence of a minimum in the intermediate region. The position of that minimum
\begin{equation}
b^{\mathrm{grav}}_{0}=\frac{\sqrt{L\lambda}}{(24\pi^2)^{1/4}}\label{eqn:b0opt}\,,
\end{equation}
can be obtained from the analysis of Eq.~(\ref{eqn:freqstab}).
At this \corr{optimum beam waist}, the \corr{phase variations} with respect to gravitational perturbations are
\begin{equation}
\corr{\sqrt{\langle \delta\phi^2 \rangle}_{\mathrm{min}} \!=\!  \left(\frac{2\pi^2}{27}\right)^{1/4}\!\!\!\frac{g L^{3/2}}{ c^2\lambda^{1/2}}= 2^{5/2}\pi^2\,\frac{g\lambda}{c^2}\!\left(\!\frac{b_{0}^{\mathrm{grav}}}{\lambda}\right)^3}\label{eqn:dfopt}.
\end{equation}
As seen from this expression, \corr{$\sqrt{\langle \delta\phi^2 \rangle}_{\mathrm{min}}$}, \corr{for fixed $\lambda$}, depends on the third power of optimum beam waist. In order to visualize this dependence, it is shown as a straight line in Fig.~\ref{FIG07} by taking $b_0=b_{0}^{\mathrm{grav}}$. As \corr{there also} can be seen, the optimum beam waist ranges between 100 $\mu$m and 300 $\mu$m, which fits very well to present experimental configurations \cite{Kessler2012, Matei2017, Keller2014}.
%

\corr{The expression (\ref{eqn:freqstab}) for the phase variations can be used to quantify the impact of Earth's gravity on a cavity system, that is calibrated with respect to the cavity symmetry axis.
It also helps to estimate the general relativistic effect on the horizontal cavity system, when it was pre-calibrated in an other orientation with respect to, or even without gravity.
Moreover, this can have qualitative consequences for cavity properties as the contrast of the output signal and its capability for frequency stabilization.
In the considered scenario, the phase fronts of the Gaussian beam in every round trip hit the plane cavity mirror under different angles, as analyzed in Sec. \ref{sec:III2}. Therefore,  a fraction of light interferes destructively and does not contribute to the cavity output intensity.
As a consequence the maximum output intensity and the contrast of the cavity output signal are reduced.  Due to the relations between the contrast of the cavity output signal and the intensity enhancement at the cavity resonance \cite{Ismail2016}, this also indicates that the effect of gravitational light deflection may influence frequency stabilization techniques, that make use of the cavity resonance curve \cite{Black2001}.}
%
%
%

\section{Summary and Conclusion} \label{sec:summary}
In this article, we present a theoretical framework to describe the propagation of light in the presence of a homogeneous gravitational field. In particular, we derived the wave equation for the vector potential of the electromagnetic field in Rindler spacetime, that accounts for leading order gravitational corrections. The wave equation is used to obtain a gravitationally modified Gaussian beam, which can be utilized to model the propagation of light in Earth-based laser experiments. As a specific setup of such an experiment, \corr{a beam, that enters a hemispherical cavity along the cavity symmetry axis}, is considered in detail. Our theory is applied to describe the round trips of light in the cavity, and the \corr{resulting} gravitational perturbations in such a device. In particular we found that these perturbations lead to \corr{phase variations at the plane mirror, that is commonly used as the cavity output.} 
Detailed calculations were performed to analyze the \corr{gravitationally induced phase variations} for a wide range of cavity settings, characterized by the beam waists $b_0$, the cavity lengths $L$ and wavelengths $\lambda$.
Special attention was paid to the effect on the currently most stable cavities with typical resonator lengths of $L=$ 21 cm, 30 cm and 50 cm, which are used to stabilize high precision instruments like optical atomic clocks and gravitational wave detectors. 
Based on our calculations
\corr{and due to the strong interconnections between phase and frequency in an optical cavity, we emphasize that gravitationally induced phase variations may have implications for cavity calibration procedures, contrast optimization and cavity-based frequency stabilization techniques.}

%
%
%

\acknowledgements{The authors would like to thank Marcel Reginatto for helpful discussions. We acknowledge the support by the Deutsche Forschungsgemeinschaft (DFG, German Research Foundation) under Germanys Excellence Strategy - EXC-2123 QuantumFrontiers - 390837967.}
%


\begin{thebibliography}{12}
	
	\bibitem{Haensch2006} T. W. Haensch, \emph{Nobel Lecture: Passion for precision}, Rev. Mod. Phys. {\bf 78}, 1297 (2006).
	
	\bibitem{Ludlow2015} A. D. Ludlow, M. M. Boyd, J. Ye, E. Peik, and P. O. Schmidt, \emph{Optical atomic clocks}, Rev. Mod. Phys. {\bf 87}, 637 (2015).
	
	\bibitem{Kessler2012} T. Kessler, C. Hagemann, C. Grebing, T. Legero, U. Sterr, F. Riehle, M. Martin, L. Chen, and J. Ye , \emph{A sub-40-mHz-linewidth laser based on a silicon single-crystal optical cavity}, Nature Photonics {\bf 6}, 687 (2012).
	
	\bibitem{Ushijima2015} I. Ushijima, M. Takamoto, M. Das, T. Ohkubo, and H. Katori, \emph{Cryogenic optical lattice clocks }, Nat. Photonics {\bf 9}, 185 (2015).
	
	\bibitem{Huntemann2016} N. Huntemann, C. Sanner, B. Lipphardt, C. Tamm, and E. Peik, \emph{Single-Ion Atomic Clock with $3\times10^{-18}$ Systematic Uncertainty}, Phys. Rev. Lett. {\bf 116}, 063001 (2016).
	
	\bibitem{Peters2001} A.~Peters, K.~Y.~Chung, and S.~Chu, \emph{High-precision gravity measurements using atom interferometry}, Metrologica {\bf 38} 25 (2001).
	
	\bibitem{Hogan2016} J. M. Hogan and M. A. Kasevich, \emph{Atom-interferometric gravitational-wave detection using heterodyne laser links}, Phys. Rev. A {\bf 94}, 033632 (2016).
	
	\bibitem{Kolkowitz2016} S. Kolkowitz, I. Pikovski, N. Langellier, M. D. Lukin, R. L. Walsworth, and J. Ye, \emph{Gravitational wave detection with optical lattice atomic clocks}, Phys. Rev. D {\bf 94}, 124043 (2016).
	
	\bibitem{Haensch2004} 
	M.~Fischer,	N.~Kolachevsky,
	M.~Zimmermann, R.~Holzwarth, T.~Udem, T.~W.~Hansch, M.~Abgrall, J.~Grunert, I.~Maksimovic, S.~Bize, H.~Marion, F.~P.~DosSantos, P.~Lemonde, G.~Santarelli, P.~Laurent, A.~Clairon, C.~Salomon, M.~Haas, U.~D.~Jentschura, and C.~H.~Keitel,
	\emph{New Limits on the Drift of Fundamental Constants from Laboratory Measurements}, Phys. Rev. Lett. {\bf 92}, 230802 (2004).
	
	\bibitem{openquest} B.~Sathyaprakash, et al., \emph{Scientific objectives of Einstein Telescope}, Classical Quantum Gravity {\bf 29}, 124013 (2012).  
	
	\bibitem{lisa} J.~Baker, et al., \emph{The Laser Interferometer Space Antenna: Unveiling the Millihertz Gravitational Wave Sky}, arXiv:1907.06482 (2019).
	
	\bibitem{Stadnik2015} \corr{Y. V. Stadnik, and V. V. Flambaum, 
	\emph{Searching for Dark Matter and Variation of Fundamental Constants with Laser and Maser Interferometry}, 
	Phys. Rev. Lett. {\bf 114}, 161301 (2015).}
	
	\bibitem{Stadnik2016} \corr{Y. V. Stadnik, and V. V. Flambaum, 
	\emph{Enhanced effects of variation of the fundamental constants in	laser interferometers and application to dark-matter detection}, 
	Phys. Rev. A {\bf 93}, 063630 (2016).}	
	
	\bibitem{Derevianko2014} A. Derevianko and M. Pospelov, \emph{Hunting for topological dark matter with atomic clocks}, Nat. Phys. {\bf 10}, 933 (2014).
	
	\bibitem{Geraci2019} A.~A.~Geraci, C.~Bradley, D.~Gao, J.~Weinstein, and A.~Derevianko, \emph{Searching for Ultralight Dark Matter with Optical Cavities} Phys. Rev. Lett. {\bf 123}, 031304 (2019)
	
	\bibitem{Matei2017} D. G. Matei, et al., \emph{1.5 $\mu$m Lasers with Sub-10 mHz Linewidth}, Phys. Rev. Lett. {\bf 118}, 263202 (2017).
	
	\bibitem{Cole2016} G. D. Cole, W. Zhang, B. J. Bjork, D. Follman, P. Heu, C. Deutsch, L. Sonderhouse, J. Robinson, C. Franz, A. Alexandrovski, M. Notcutt, O.H. Heckl, J. Ye, and M. Aspelmeyer, \emph{High-performance near- and mid-infrared crystalline coatings}, Optica {\bf 3}, 647 (2016).
	
	\bibitem{Dickmann2018-2} J. Dickmann and S. Kroker, \emph{Highly reflective low-noise etalon-based meta-mirror},
	Phys. Rev. D {\bf 98}, 082003 (2018).
	
	\bibitem{quantumnoise} M.~T.~Jaeckel, and S. Reynaud, \emph{Quantum Limits in Interferometric Measurements}, EPL {\bf 13}, 301 (1990). 
	
	\bibitem{thermooptic} M.~Evans, S.~Ballmer, M.~Fejer, P.~Fritschel, G.~Harry, and G.~Ogin, \emph{Thermo-optic noise in coated mirrors for high-precision optical measurements}, Phys. Rev. D {\bf 78}, 102003 (2008).
		
	\bibitem{Einstein16} A.~Einstein, 
	\emph{The Foundation of the General Theory of Relativity},  
	Annalen der Physik {\bf 49}, 769 (1916).
	
	\bibitem{Edd19}  F.~W.~Dyson, A.~S.~Eddington, and C.~Davidson, 
	\emph{A determination of the deflection of light by the sun's gravitational field, from observations made at the total eclipse of May 29}, 
	Philosophical Transactions of the Royal Society of London. Series A, {\bf 220}, 291 (1920).
	
	\bibitem{Raetzel2018} \corr{D.~Rätzel, F.~Schneiter, D.~Braun, T.~Bravo, R.~Howl,
	M.~P.~E.~Lock, and I.~Fuentes,
	\emph{Frequency spectrum of an optical resonator in a curved spacetime},
	New J. Phys. {\bf 20}, 053046 (2018).} 
	
	\bibitem{Rich19}  M.~Richard, 
	\emph{Free-fall of photons in a planar optical cavity}, 
	J. Phys. Commun. {\bf 3}, 045007 (2019).
	
	\bibitem{Ulbricht2020} S.~Ulbricht, J.~Dickmann, R.~A.~Müller, S.~Kroker, A.~Surzhykov, \emph{Gravitational light deflection in Earth-based laser cavity experiments}, Phys. Rev. D {\bf 101}, 121501(R) (2020). 
	
	\bibitem{Rind60} W.~Rindler, \emph{Hyperbolic Motion in Curved Space Time}, Phys. Rev. Lett. {\bf 119}, 2082 (1960).
	
	\bibitem{Rind66} W.~Rindler, \emph{Kruskal space and the uniformly accelerated frame}, Am. J. Phys. {\bf 34}, 1174 (1966).
	
	\bibitem{Wald} R.~M.~Wald, \emph{General Relativity}, University of Chicago Press, Chicago (1984).
	
	\bibitem{Carrol} S.~M.~Carrol, \emph{An Introduction to General Relativity - Spacetime and Geometry}, Addison Wesley, San Francisco (2004).
	
	\bibitem{Remark2} Due to  $\nabla_{\mu}A^{\mu}=0$, the scalar potential is determined by $
	\Phi[\vec{A}]=-\frac{i c^2}{\omega_0}\left(1+\frac{g z}{c^2}\right) \vec{\nabla} \!\cdot\!\left[\left(1+\frac{g z}{c^2}\right) \vec{A}\right]$\,.
	
	\bibitem{Hecht2002} E.~Hecht, \emph{Optics}, Vol. 4, Addison Wesley, San Francisco (2002).
	
	\bibitem{Hunt81} P.~M.~Hunt, \emph{A continuum basis of airy functions: Matrix elements and a test calculation},  Mol. Phys., {\bf 44}, 653-663 (1981).
	
	\bibitem{Vallee} O.~Vall\'ee, and M.~Soares, \emph{Airy Functions and Applications to Physics}, Imperial College Press, London (2004).
	
	\bibitem{horizontalcavity} J. Davila-Rodriguez, F.~N.~Baynes, A.~D.~Ludlow, T.~M.~Fortier, H.~Leopardi, S.~A.~Diddams, and F.~Quinlan, \emph{Compact, thermal-noise-limited reference cavity for ultra-low-noise microwave generation} , Optics letters {\bf 42}, 1277 (2017).
	
	\bibitem{Saulson} P.~R.~Saulson, \emph{Fundamentals of Interferometric Gravitational Wave Detectors}, World Scientific, Singapore (1994).
	
	\bibitem{Remark1}
	Here, only the $z'$-integral differs from an ordinary Fourier transformation. It reads
	\begin{eqnarray}
	\hspace{2em}\int\limits_{-\infty}^{\infty}\!\e^{-\frac{z'^2}{2b_0^2}}
	\Ai\left[ -\left(\frac{k_z}{\delta k_z}\right)^2\!\corr{+}\delta k_z z'  \right]\d {z'} \nonumber\,,
	\end{eqnarray}
	where the damping term of the basis function and the redshift factor $\e^{\corr{-}\gamma z'}=(1\corr{-}gz'/2c^2)+\mathcal{O}(\epsilon^2)$ of the boundary condition cancel the redshift factor of the integral measure $(1\corr{+}2\gamma z')\d z'$ to order $\mathcal{O}(\epsilon^2)$.
	
	\bibitem{Dirichlet}  D.~Zwillinger, \emph{Handbook of Differential Equations}, Academic Press, San Diego (1989).
	
	\bibitem{Meschede} \toproof{D. Meschede, \emph{Optik, Licht und Laser}, Springer-Verlag (2009).}
	
	\bibitem{Didier2019} A.~Didier, S.~Ignatovich, E.~Benkler, M.~Okhapkin, and T.~E.~Mehlstäubler, \emph{946-nm Nd:YAG digital-locked laser at $1.1\times 10^{-16}$ in 1 s and transfer-locked to a cryogenic silicon cavity}, Opt. Lett. {\bf 44}, 1781 (2019).
	

	\bibitem{Keller2014} J. Keller, et al., \emph{Simple vibration-insensitive cavity for laser stabilization at the $10^{-16}$ level}, Applied Physics B {\bf 116.1}, 203 (2014).

	\bibitem{Nakagawa1994} K.~Nakagawa, T.~Katsuda, A.~S.~Shelkovnikov, M.~de~Labachelerie, and M.~Ohtsu, \emph{Highly sensitive detection of molecular absorption using a high finesse optical cavity}, Opt. Commun.  {\bf 107}, 369 (1994).
	
	\bibitem{Ismail2016} N.~Ismail, C.~C.~Kores, D.~Geskus, and M.~Pollnau, \emph{Fabry-Pérot resonator: spectral line shapes, generic and related Airy distributions, linewidths, finesses, and performance at low or frequency-dependent reflectivity},
	Opt. Express {\bf 24}, 16366 (2016).
	
	\bibitem{Dawkins2007} S.~T.~Dawkins, J.~J.~McFerran, and A.~N.~Luiten, \emph{Considerations on the measurement of the stability of oscillators with frequency counters}, 
	IEEE Transactions on Ultrasonics, Ferroelectrics, and Frequency Control {\bf 54}, 918 (2007). 
	
	\bibitem{Ludlow2007} A.~D.~Ludlow, X.~Huang, M.~Notcutt, T.~Zanon-Willette, S.~M.~Foreman, M.~M.~Boyd, S.~Blatt, and J.~Ye, \emph{Compact, thermal-noise-limited optical cavity for diode laser stabilization at $1\times10^{-15}$}, Opt. Lett. {\bf 32}, 641 (2007). 
    
    \bibitem{Black2001}
    \corr{E.~D.~Black,
    \emph{An introduction to Pound–Drever–Hall laser frequency stabilization},
     Am. J. Phys. {\bf 69}, 79 (2001).}
	
\end{thebibliography}
\end{document}